\documentclass[conference]{IEEEtran}
\usepackage{cite}
\usepackage{amsmath,amssymb,amsfonts}
\usepackage{algorithmic}
\usepackage{graphicx}
\usepackage{textcomp}
\usepackage{xcolor}
\usepackage{caption}
\usepackage{subcaption}
\usepackage{float}
\makeatletter
\newcommand{\linebreakand}{%
  \end{@IEEEauthorhalign}
  \hfill\mbox{}\par
  \mbox{}\hfill\begin{@IEEEauthorhalign}
}

\def\BibTeX{{\rm B\kern-.05em{\sc i\kern-.025em b}\kern-.08em
    T\kern-.1667em\lower.7ex\hbox{E}\kern-.125emX}}
\begin{document}

\title{FPGA based Implementation of \\Frequency and Phase Matching Technique \\for Grid Tied Applications
{\footnotesize \textsuperscript{}}
\thanks{}
}

\author{\IEEEauthorblockN{Uzair Nadeem}
\IEEEauthorblockA{\textit{Department of Electrical Engineering, College of EME} \\
\textit{National University of Sciences and Technology (NUST)}\\
Rawalpindi, Pakistan \\
uzairnadeem92@outlook.com}

\and
\IEEEauthorblockN{Muhammad Shahzaib Atif}
\IEEEauthorblockA{\textit{Department of Electrical Engineering, College of EME} \\
\textit{National University of Sciences and Technology (NUST)}\\
Rawalpindi, Pakistan \\
shahzaibmsa@gmail.com}

\linebreakand
\IEEEauthorblockN{Rizwan Ahmed}
\IEEEauthorblockA{\textit{Department of Electrical Engineering, College of EME} \\
\textit{National University of Sciences and Technology (NUST)}\\
Rawalpindi, Pakistan \\
rizwanahmed32@ee.ceme.edu.pk}
\and

\IEEEauthorblockN{Hassan Touqeer}
\IEEEauthorblockA{\textit{Department of Electrical Engineering, College of EME} \\
\textit{National University of Sciences and Technology (NUST)}\\
Rawalpindi, Pakistan \\
hassantouqeer32@ee.ceme.edu.pk}
\linebreakand

\IEEEauthorblockN{Hamood Ur Rahman Khawaja}
\IEEEauthorblockA{\textit{Department of Electrical Engineering, College of EME} \\
\textit{National University of Sciences and Technology (NUST)}\\
Rawalpindi, Pakistan \\
hamood@ceme.nust.edu.pk}
}

\maketitle

\begin{abstract}
A grid tied inverter converts DC voltage into AC voltage, while synchronizing it with the supply line phase and frequency. This paper presents an efficient, robust, and easy-to-implement grid tie mechanism. First, the grid tie mechanism was simulated in software using LabVIEW and Multisim. Then, the whole system was practically implemented on hardware. A prototype hardware was developed to produce AC voltage from solar panels. Phase and frequency of the generated voltage were synchronized with those of a  reference sinusoidal signal. The synchronization mechanism was digitally implemented on an FPGA, which also controlled the whole system. We achieved real time frequency matching with an improved Zero Crossing Detection (ZCD) technique. Phase matching was also achieved in real time using a modified Phase Locked Loop (PLL) algorithm, which retains stability while being simpler than the general PLL algorithm. Experiments demonstrated that the proposed grid tied system reliably synchronized the phase and frequency of the voltage generated by the implemented hardware with those of the reference grid voltage.
\end{abstract}

\section{Introduction}\label{AA}
The ever increasing energy requirement is one of the biggest challenges faced by the modern world. The depletion of fossil fuels for energy generation is one of the major concerns of the modern age. This has led to a vast exploration of various methods employed for the generation of electricity from renewable sources. The use of solar energy as (virtually) an infinite source of power seems to be the most plausible answer to this situation. Solar panels and photo-voltaic (PV) cells have become the center of attention in this entire scenario.

Solar energy can become a viable solution to the problem of energy crisis \cite{b1}. However, there are still many challenges for the effective utilization of solar energy. One of the challenges is the efficient conversion of Direct Current (DC) electricity harnessed from solar panels to Alternating Current (AC) electricity.  Many different inverter designs have been proposed so far for the conversion of DC to AC \cite{b2,b3}. The next major challenge for utilizing solar energy in the national grid is the frequency and phase matching of the generated AC voltage with that of the grid. The implementation of grid tie mechanism on smart local devices is a challenging task \cite{b4}.

This paper deals with the implementation of frequency and phase matching algorithms on a prototype hardware. The hardware consists of solar panels, a DC-DC converter, a full H-bridge inverter, an analog filter, and an FPGA \cite{b5}. Output of the solar panels was passed through a DC-DC converter, and then fed to a full H-bridge inverter. Sinusoidal Pulse Width Modulation (SPWM) was used to drive the inverter circuit \cite{b6}. Inverter was followed by an analog filter to produce a sinusoidal output voltage. Phase and frequency of the generated output voltage were synchronized with that of the grid's voltage. The synchronization was controlled using the FPGA of the General Purpose Inverter Controller (GPIC) of National Instruments (NI). 

Rest of the paper is organized as follows: Section \ref{BB} discusses the overall mechanism of the grid tied system. Section \ref{CC} describes the simulations and the implementation of the proposed frequency matching technique. Details of the phase matching and PLL (Phase Locked Loop) are presented in Section \ref{DD}. Finally, the paper is concluded in Section \ref{EE}.

\section{Description of the Grid Tied Operation}\label{BB}

In order to grid tie the produced AC voltage, frequency and phase of the generated output voltage are to be matched with those of the grid's voltage. We used Zero Crossing Detection (ZCD) technique to match the frequency of the generated signal with the frequency of the AC signal obtained from the national grid. Phase of the generated signal was also matched with the phase of the grid's signal using Phase locked Loop (PLL), which was vital in achieving phase synchronization. Both ZCD and PLL were implemented digitally on the FPGA of GPIC. Appropriately scaled feedback from the generated voltage and the grid's voltage were used as inputs to the FPGA. Based on the values of these feedback signals, FPGA controlled the SPWM fed to the inverter, which in turn was responsible for performing frequency and phase matching tasks. 

The disadvantage of using normal square pulse train to operate an H-bridge is that the resulting sinusoidal waveform contains higher frequency harmonics, even after filtration. Therefore, SPWM was used as an input to the inverter. SPWM greatly reduced the third and higher order harmonics in the output voltage of the inverter. FPGA of GPIC acted as an on-board computer to sense the current AC grid waveform and produced voltage to correspond with grid. 

\begin{figure}[tbp]
\centering
\includegraphics[width=\linewidth]{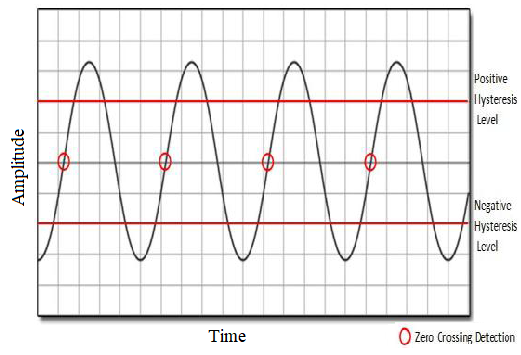}
\caption{A representation of the improved Zero Crossing Detection (ZCD) technique and the hysteresis levels.}
\label{fig1}
\end{figure}

\section{Frequency Matching}\label{CC}

Frequency matching is a process in which frequency of the output voltage of the grid tied system is matched with the frequency of the desired signal (grid's electricity signal). Encouraging results were obtained for frequency matching using an improved ZCD technique. The following subsections discuss the ZCD technique and its practical implementation.

\subsection{Implementation of ZCD Technique}
As discussed earlier, frequency of the inverter output should match with the frequency of the reference voltage, i.e., the national grid, in order for the grid tied system to work properly. As a first step of frequency matching, the reference voltage was fed to the FPGA through an Analog to Digital Converter (ADC), after scaling it down to the desired level. Input frequency was determined by detecting zero crossings using FPGA. Zero crossings are points where sinusoidal signal crosses the zero voltage level. After determining the input frequency, an SPWM of the same frequency was generated by the FPGA. This SPWM was then utilized to operate the inverter.

\subsection{Improvement of ZCD Technique}
ZCD technique generally does not perform reliably in noisy environment. Hysteresis technique was used to overcome this problem. In this technique, zero crossings are detected only on either increasing or decreasing side. In our approach, zero crossings were detected on the increasing side, i.e., while the voltage was going from negative to positive. FPGA detected frequency by detecting only two zero crossings. Once a zero crossing was detected, two steps were performed in succession to validate it, before detection of another zero crossing. Firstly, voltage level must increase above a threshold, called the positive hysteresis level. Secondly, voltage level must decrease below a threshold, called the negative hysteresis level. Next, zero crossing was detected only after the successful detection of these two steps in succession. Any possible distortion caused by noise in the detection of the grid frequency was eliminated by appropriately determining hysteresis level. Fig.~\ref{fig1} shows a representation of the zero crossing detection on the increasing side and the hysteresis levels.

\begin{figure}[b!]
\centering
\includegraphics[width=\linewidth]{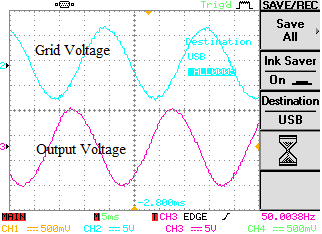}
\caption{Output screen of oscilloscope for frequency matching at 50 Hz.}
\label{fig2}
\end{figure}
\begin{figure}[tbp]
\centering
\includegraphics[width=\linewidth]{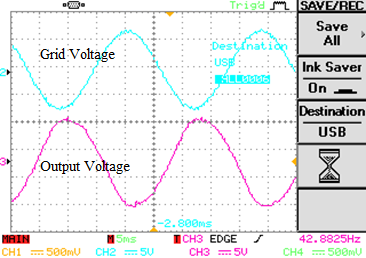}
\caption{Output screen of oscilloscope for frequency matching at 42.88 Hz.}
\label{fig3}
\end{figure}

\subsection{Software Simulations}
Before the hardware implementation of the frequency matching, the algorithm was first tested through software simulations. LabVIEW and Multisim softwares were used to test the algorithm with various levels of noise. Through the improved ZCD technique, the frequency of the generated output voltage was successfully matched with that of the reference voltage. 

\begin{figure*}[hbt!]
\centering
\includegraphics[width=0.7\linewidth]{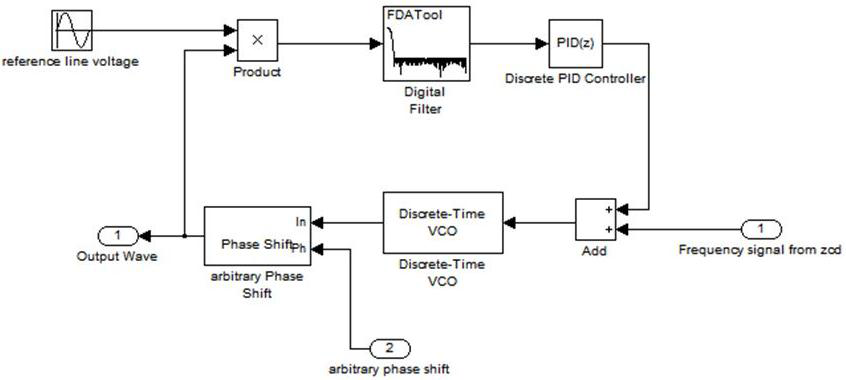}
\caption{A block diagram of the proposed digitally implemented Phase Locked Loop (PLL) algorithm.}
\label{fig4}
\end{figure*}
\subsection{Hardware Test}
The nominal frequency of the national grid's voltage is 50 Hz. In a grid tied system, grid's voltage is the reference voltage whose frequency and phase are to be matched by the system. For testing, we used a sinusoidal signal produced from a function generator as a reference voltage (in place of the grid's voltage). Frequency matching test was then carried out on a number of frequencies, in the range of 35 Hz to 65 Hz. As frequency of the reference voltage was varied, frequency of the generated output voltage of grid tied inverter also varied accordingly in real time.

\subsection{Results of Frequency Matching Tests}
The frequency matching was reliably achieved during both simulations and hardware implementations. Fig.~\ref{fig2} and Fig.~\ref{fig3} show outputs of the oscilloscope for the frequency matching tests carried out in the lab environment at 50 Hz and 42.88 Hz, respectively. Frequency matching was achieved up to two decimal places. The maximum settling time is two cycles of the reference input wave. For instance, the settling time for a nominal 50 Hz signal is 40 ms. It was not possible to determine the error and precision for more than two decimal places due to the limitation of the testing equipment. Note that the signals are out of phase in Fig.~\ref{fig2} and Fig.~\ref{fig3} as only frequency matching was performed at this stage.

\section{Phase Matching}\label{DD}
Phase Locked Loop (PLL) is a closed loop control system that generates an output signal whose phase is related to the phase of the input signal. 

\subsection{Implementation of PLL}
A PLL consists of a voltage controlled oscillator (VCO), phase detector and a low pass filter \cite{b7}. We implemented PLL digitally (digital phase locked loop) on the FPGA of NI GPIC using LabVIEW. Instead of voltage controller oscillator, a numerically controlled oscillator (NCO) was programmed on FPGA. Phase detector detects and eliminates the phase differences between the reference input signal and the generated voltage. Then, the output signal is again fed to the input and the process continues in real time. 

PLL loop filter (usually a low pass filter), has two distinct functions. Firstly, its main function is to determine the loop dynamics (stability). This is how the loop responds to disturbances. Disturbances may involve changes in the reference frequency, the feedback divider and the initial mismatch. Secondly, it limits the amount of reference frequency energy (ripple) appearing at the output of phase detector, which is applied to the control input of NCO.

\subsection{Phase Matching Algorithm}
Fig.~\ref{fig4} shows implementation of the proposed PLL algorithm. This algorithm is simpler and more efficient to implement on FPGA than the standard PLL algorithm. Analog-to-Digital converters (ADCs) were used to feed the scaled down versions of the generated voltage and the grid voltage to the FPGA. Product of line voltage (reference voltage) and the inverter final output voltage (output wave) was fed to a low pass digital filter. Low pass filtering removes high frequency components. Only the DC component, called the error signal, is extracted from the output of the low pass filter for the next stage. A digitally implemented Proportional-Integral-Derivative (PID) controller was used to improve the loop performance. The DC component from the output of the digital filter was fed as the error signal to the PID controller. Output of the PID controller was then added to the frequency measurement signal from the ZCD part (Section~\ref{CC}), which was then fed to the NCO (Discrete Time VCO). Phase shift block in Fig.~\ref{fig4} represents any arbitrary phase shift which may occur due to filter, transformer or load effects. The PID controller was designed to settle the process variable to 0.5 on the basis of the following equations:
\begin{multline}\label{eq1}
V_1*V_0=
\frac{1-\cos(4\pi ft)}{2} * \cos\theta  \\
+ \frac{1-\sin(4\pi ft)}{2} * \sin\theta
\end{multline}
\begin{multline}\label{eq2}
V_1*V_0=
\frac{\cos\theta}{2} - \frac{\cos(4\pi ft)}{2} * \cos\theta  \\
+ \frac{\sin(4\pi ft)}{2} * \sin\theta
\end{multline}
where V\textsubscript{1} =  $\sin(2\pi ft$), V\textsubscript{0} =  $\sin(2\pi ft + \theta$)

\begin{figure}[b!]
\centering

\begin{subfigure}[htbp]{\linewidth}
\centering
\includegraphics[width=\linewidth]{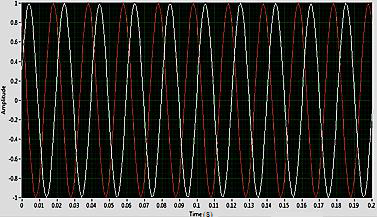}
\caption{The simulated reference and the generated voltage waveforms before phase matching (Best viewed in color).}
\label{fig5a}
\end{subfigure}

\begin{subfigure}[htbp]{\linewidth}
\centering
\includegraphics[width=\linewidth]{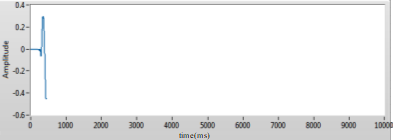}
\caption{Error signal (process variable) to the PID controller at the start of phase matching process (simulation).}
\label{fig5b}
\end{subfigure}

\caption{Waveforms of grid voltage, generated voltage and the error signal (process variable) to the PID controller at the start of simulations for phase matching.}
\label{fig5}
\end{figure}

\begin{figure}[b!]
\centering

\begin{subfigure}[htbp]{\linewidth}
\centering
\includegraphics[width=\linewidth]{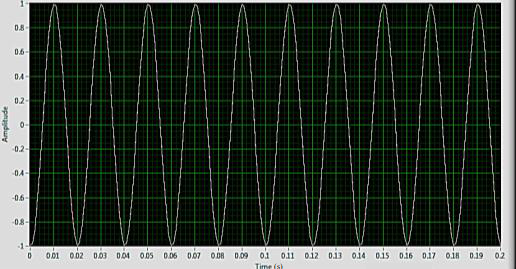}
\caption{The simulated reference and the generated voltage waveforms after phase matching (Best viewed in color).}
\label{fig6a}
\end{subfigure}

\begin{subfigure}[htbp]{\linewidth}
\centering
\includegraphics[width=\linewidth]{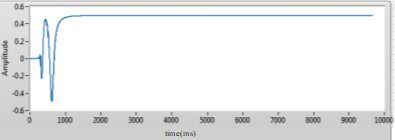}
\caption{Error signal (Process variable) to the PID controller showing phase matching process (simulation). The error signal has settled at the expected value of 0.5 (normalized). See Section \ref{DD} for detail.}
\label{fig6b}
\end{subfigure}

\caption{Waveforms of grid voltage, generated voltage and the error signal (process variable) to the PID controller after phase matching has been achieved in simulations.}
\label{fig6}
\end{figure}
\subsection{Results of Phase Matching Tests}
Once the phase of the output wave is matched with the phase of the input wave, the relative phase difference should be zero. In order to achieve frequency and phase matching, the PID controller attempts to settle the error signal (input process variable) to 0.5 (normalized). 

\subsubsection{Simulations}

Before hardware implementation, we performed simulations of the proposed PLL algorithm using LabVIEW. Fig.~\ref{fig5} and Fig.~\ref{fig6} show the grid voltage, generated voltage and the error signal (process variable) to the PID controller during simulations.

Fig.~\ref{fig5a} shows two sinusoidal waveforms before phase matching. The waveform with a positive cycle first (the one with a lighter shade) represents grid voltage, whereas, the waveform starting with a negative cycle represents the output voltage of the prototype grid tied system. Inverter output voltage tries to match its phase with that of  the grid's voltage from the very start. Fig.~\ref{fig5b} shows variations in input signal to the PID controller at the start of phase matching. 

Fig.~\ref{fig6a} shows the voltage from the grid and the voltage from the grid tied inverter, after phase matching has been achieved. The error signal to the PID controller settled at the desired value of 0.5, as shown in Fig.~\ref{fig6b}. 
The proposed PLL algorithm took less than one second to achieve phase synchronization and then tracked the phase of the grid voltage continuously in real time.

\subsubsection{Hardware Results}

\begin{figure}[t!]
\centering
\includegraphics[width=\linewidth]{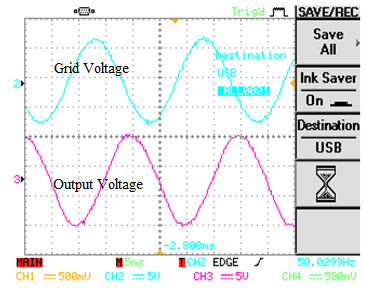}
\caption{Output screen of oscilloscope showing the reference and the generated waveforms before phase matching.}
\label{fig7}
\end{figure}

\begin{figure}[tbp]
\centering
\includegraphics[width=\linewidth]{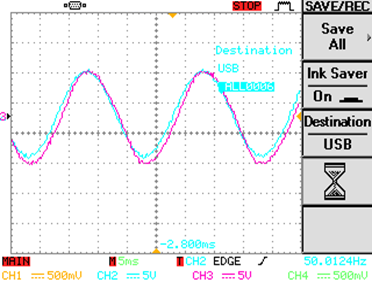}
\caption{Output screen of oscilloscope showing the reference and the generated waveforms after phase matching.}
\label{fig8}
\end{figure}

For hardware testing, we used sinusoidal voltage signal from a function generator as the reference voltage. Fig.~\ref{fig7} and Fig.~\ref{fig8} show results obtained after practical implementation of phase matching. Fig.~\ref{fig7} shows the reference and the generated voltage waveforms before phase matching, while Fig.~\ref{fig8} shows waveforms after phase matching has been achieved. Slight difference in waveforms in Fig.~\ref{fig8} is due to the fact that reference waveform has no third harmonic (as it is generated from a function generator). However, waveform generated by the grid tied system has a small non-zero third harmonic (as it is the result of an SPWM inverter followed by a filter). The grid tied inverter output voltage is always kept slightly higher than the grid's voltage, as is also obvious in Fig.~\ref{fig8}. PLL tracked the phase of the reference voltage and then synchronized the phase of the generated voltage with it.

\section{Conclusion}\label{EE}
This paper has presented the procedure and results for the simulation and implementation of an efficient and robust mechanism for grid tying a system. We reliably achieved the main objectives of real time frequency and phase matching. We used Zero Crossing Detection (ZCD) technique for frequency matching. Effect of noise was canceled with the help of hysteresis levels for an improved ZCD implementation. Experiments in the frequency range of 35 Hz to 65 Hz demonstrated that the frequency of the generated voltage was successfully matched with that of the reference voltage up to two decimal places with a maximum settling time equal to that of two waveforms of the reference signal. A modified Phase Locked Loop (PLL) was used for phase matching of the two waveforms. The PLL was tuned for a set point of 0.5 (normalized). The modified PLL took less than one second start-up time, i.e. to synchronize the phase at start, and then was able to track the phase of the reference wave in real time. The techniques mentioned in this paper can easily be implemented on an embedded hardware for grid tying purposes.
\section*{Acknowledgment}
This work is partially supported by grant and hardware from National Instruments Corporation NI (Pakistan Office). We are grateful for their support in carrying out this research work.

\end{document}